\documentstyle[pra,aps,multicol,epsfig]{revtex}

\input epsf.tex

\newcommand{\be}{\begin{equation}}
\newcommand{\ee}{\end{equation}}
\newcommand{\ba}{\begin{eqnarray}}
\newcommand{\ea}{\end{eqnarray}}
\newcommand{\ban}{\begin{eqnarray*}}
\newcommand{\ean}{\end{eqnarray*}}

\newcommand{\ket}[1]{\mbox{$ | #1 \rangle $}}
\newcommand{\bra}[1]{\mbox{$ \langle #1 | $}}

\newcommand{\si}{\sigma}
\newcommand{\demi}{\frac{1}{2}}
\newcommand{\compl}{\begin{picture}(8,8)\put(0,0){C}\put(3,0.3){\line(0,1){7}}\end{picture}}

\newcommand{\one}{\leavevmode\hbox{\small1\normalsize\kern-.33em1}}

\begin{document}

\title{Thermalizing Quantum Machines: Dissipation and Entanglement}
\author{Valerio Scarani$^1$, M\'ario Ziman$^2$,
Peter \v{S}telmachovi\v{c}$^2$, Nicolas Gisin$^1$, Vladim\'{\i}r
Bu\v{z}ek$^{2,3}$ }
\address{$^1$ Group of Applied Physics, University of Geneva, 20, rue de
l'Ecole-de-M\'edecine, CH-1211 Geneva 4, Switzerland
\\
$^2$ Research Center for Quantum Information, Slovak Academy of Sciences, D\'ubravsk\'a cesta 9, 842 28 Bratislava, Slovakia\\
$^3$ Faculty of Informatics, Masaryk University, Botanick\'a 68a, 602 00 Brno, Czech Republic\\
}
\maketitle

\begin{abstract}
We study the relaxation of a quantum system towards the thermal
equilibrium using tools developed within the context of quantum
information theory. We consider a model in which the system is a
qubit, and reaches equilibrium after several successive two-qubit
interactions (thermalizing machines) with qubits of a reservoir.
We characterize completely the family of thermalizing machines.
The model shows a tight link between dissipation, fluctuations,
and the maximal entanglement that can be generated by the
machines. The interplay of quantum and classical information
processes that give rise to practical irreversibility is
discussed.
\end{abstract}

\begin{multicols}{2}

The hypothesis of the quantum appeared suddenly in physics as an
offspring of thermodynamics, due to the work of Planck on the
blackbody radiation. In its early days however, the new theory
developed rather as a form of reversible mechanics. One century
after Planck's intuition, the link between quantum mechanics (QM)
and thermodynamics has been discussed by several scientists, and
is still an actual field of research \cite{mahler}. In parallel to
fundamental issues, the concept of {\em quantum machines} has
arisen recently in the field of quantum information processing
\cite{machines}. Looking back again to history, we see that
thermodynamics was born to describe engines. It is thus natural to
ask whether there is a "thermodynamics" of quantum machines, and
whether the modern standpoint of quantum information can cast some
new light on the foundations of thermodynamics. After some
pioneering works \cite{sanduhor}, these ideas have stimulated many
investigations in the last months \cite{carnot}.

In this Letter, we focus on the process of {\em thermalization},
that is, the relaxation towards the thermal equilibrium of a
system in contact with a huge reservoir (bath). More precisely,
let $\rho_B$ be the thermal state of the bath \cite{notehypo},
$\rho$ a generic state of the system, and $\rho^{e}$ the state of
the system at thermal equilibrium. A thermalization process is
defined by these two requirements: (I) The state
$\rho^{e}\otimes\rho_B$ is stationary; (II) If the system is
prepared in a state $\rho\neq \rho^{e}$, at the end of the process
we have a total state $\rho_{SB}$ such that
$\mbox{Tr}_B[\rho_{SB}] \simeq \rho^{e}$ and
$\mbox{Tr}_S[\rho_{SB}] \simeq \rho_B$, where $\mbox{Tr}_{B,S}$
are the partial trace over the bath and the system respectively.

\begin{center}
\begin{figure}
\epsfxsize=8cm \epsfbox{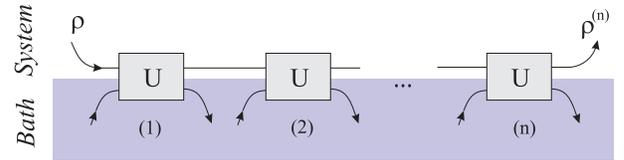} \caption{The quantum channel:
 a repeated application of a unitary $U$ (quantum machine), that
 couples the state of the system with the state of the bath.}
\label{figbath}
\end{figure}
\end{center}

Thermalizing quantum channels can be realized by letting the
system undergo interactions with the bath that are localized in
time. Such models, known as {\em collision models} \cite{alicki},
are admittedly rather artificial as models for dissipative
processes \cite{weiss}, but are most natural in the context of
quantum information \cite{ter}. The system passes through several
identical machines $U$ (figure \ref{figbath}), or several time
through the same machine; at each passage, it becomes entangled
(that is, it shares a part of the information encoded in the
state) with an ancilla, i.e. some degrees of freedom of the bath.
At the output of the machine, the ancilla is discarded into the
bath: the information present in the system has undergone some
degradation, that depends on the state of the bath and on the
machine.

Our main goal is to quantify the role of {\em entanglement} in
this thermalization process. Since a computable measure for
entanglement of mixed states is known only for states of two
two-dimensional quantum systems (qubits) \cite{woot}, we consider
a thermalization process in which both the system and the ancillas
are qubits. Before discussing entanglement, we give the family of
all the {\em thermalizing machines} $U$ acting on two qubits, and
a fluctuation-dissipation theorem for the thermalizing channel
that these $U$ define.

{\em The model.} We start with a description of the model:\\
(i) The system is a qubit, and the bath is a reservoir composed of
an arbitrary large number $N$ of qubits. The free hamiltonian for
the whole system is \ba H_0\,=\,H_S+H_B&=& h[S]+\sum_{i=1}^{N}h[i]
\label{hamiltonian}\ea where $h[k]$ is the operator acting as
$h=-E\si_z$ on the qubit $k$ and trivially on the other qubits.
The bath is supposed to be initially in the thermal state
$\rho_B\,=\, e^{-\beta H_B}/\mbox{Tr}\big(e^{-\beta
H_B}\big)=(\xi)^{\otimes N}$ with $\xi=e^{-\beta
h}/\mbox{Tr}\big(e^{-\beta h}\big)=\demi\big(\one+ \tanh (\beta
E)\si_z \big)$, and $\beta=\frac{1}{kT}$. Let $P_0=\ket{0}\bra{0}$
and $P_1=\ket{1}\bra{1}$ be the projectors on the eigenstates of
$\si_z$; thus \ba \xi\,=\, pP_{0}+qP_{1}\;&,&\;q=1-p
\label{statebath} \ea with $p=\demi(1+\tanh (\beta E))$. We set
$E>0$, so that $\ket{0}$ is the ground state, and $p=1$
corresponds to $T=0$.\\
(ii) The machine $U$ is a unitary operation on
$\compl^2\otimes\compl^2$. This means that at fixed time the
system interacts with just a {\em single} qubit taken out of the
bath. In our model we consider that a qubit of the bath undergoes
at most one interaction with the system. This assumption is
justified by the fact that the bath is assumed to very large (i.e.
"infinite"). Therefore the input state of the ancilla is always
$\xi$, and we write \ba \rho^{(k+1)}&=&\mbox{Tr}_B\big[
U\,(\rho^{(k)}\otimes
\xi)\,U^{\dagger}\big]\,\equiv\,T_{\xi}\,[\rho^{(k)}]\,.
\label{rhoout} \ea

{\em Thermalizing machines.} For the model just introduced, the
two requirements I and II read \cite{note2}: \ba \mbox{Req.
I:}&~~& U_z\,(\xi\otimes\xi)\,U_z^{\dagger} = \xi\otimes\xi \; \;
,\,\xi = p P_0 + q P_1\;,\forall \,p\label{req1}
\\
\mbox{Req. II:}&~~& \rho^{(n)}=T_{\xi}^n[\rho]\,\longrightarrow
\xi\;\;\forall\,\rho\,. \label{req2} \ea The subscript $z$ is
meant to remind that we allow the machine $U$ to depend on the
eigenbasis of $\xi$, that is on $h$. Conversely, we want these
requirements to hold for all $p$, that is for all temperature.

It is important to notice here the existence of an equivalence
class. Let $u(x)=P_0+e^{ix}P_1$, and suppose that $U_z$ satisfies
(\ref{req1}) and defines a channel $T_{\xi}$. Then $U'_z=
\big(\one\otimes u(\alpha)\big)\,U_z\, \big(\one\otimes
u(\beta)\big)$ satisfies (\ref{req1}) as well, and defines the
same channel $T'_{\xi}=T_{\xi}$. This is easy to see by noticing
that $u(x)\xi u(x)^{\dagger}=\xi$ for all $x$. This equivalence is
a consequence of the freedom of choosing the global phases of
$\ket{0}$ and $\ket{1}$ for qubits in the bath. Having noticed
this, we can proceed to find all the thermalizing machines.

Take first Requirement I: condition (\ref{req1}) implies that the
subspaces $P_0\otimes P_0$, $P_1\otimes P_1$ and $P_0\otimes P_1+
P_1\otimes P_0$ must be invariant under the action of $U$. In
fact, on the l.h.s. $U_z\,P_0\otimes P_0\,U_z^{\dagger}$ appears
with the weight $p^2$, $U_z\,\big(P_0\otimes P_1+ P_1\otimes
P_0\big)\,U_z^{\dagger}$ with the weight $p(1-p)$, and
$U_z\,P_1\otimes P_1\,U_z^{\dagger}$ with the weight $(1-p)^2$.
Since we want condition (\ref{req1}) to hold for all $p$, the
three subspaces must be separately invariant. This implies
$[U_z,H_0]=0$: the sum of the one-qubit energies is conserved by
the interactions. By inspection, one can see that, up to a global
phase factor, the most general unitary operation that leave these
subspaces invariant is parametrized by five angles; only three of
them are left if we choose a suitable representative element in
the equivalence class discussed above. Precisely, all unitary
operations that fulfill the condition (\ref{req1}) can be chosen
of the form \ba
\begin{array}{clcl}
U_z(\phi,\theta,\alpha):&\ket{0}\ket{0}&\longrightarrow&\ket{0}\ket{0}\\
&\ket{1}\ket{1}&\longrightarrow&\ket{1}\ket{1}\\
&\ket{0}\ket{1}&\longrightarrow&e^{i(\theta+\alpha)}\big(c\ket{0}\ket{1}+
is\ket{1}\ket{0}\big)\\
&\ket{1}\ket{0}&\longrightarrow&e^{i(\theta-\alpha)}\big(c\ket{1}
\ket{0}+ is\ket{0}\ket{1}\big)\, ,
\end{array}
\label{generalu} \ea with $c=\cos\phi$ and $s=\sin\phi$,
$\phi\in[0,\frac{\pi}{2}]$, and $\theta,\alpha\in[0,2\pi]$. We
turn now to demonstrate that almost all these machines satisfy the
condition (\ref{req2}) as well. To do this, let's write the state
of the system after $n$ steps as \ba \rho^{(n)}&=&d^{(n)}\,
P_0\,+\,(1-d^{(n)})\,
P_1\,+\,k^{(n)}\ket{0}\bra{1}\,+\,\mbox{h.c.} \label{statein} \ea
Inserting the explicit form (\ref{generalu}) for $U_z$ into
(\ref{rhoout}), we find that the effect of the map $T_\xi$ is
given by $d^{(n+1)}=d^{(n)}c^2+ps^2$ and $k^{(n+1)}=c\,\lambda
\,k^{(n)}$ with \ba \lambda&=&e^{i\alpha}\,\big(pe^{-i\theta} +
qe^{i\theta}\big)\,.\label{lambda}\ea A straightforward iteration
gives $d^{(n)}$ and $k^{(n)}$ as a function of the parameters
$d^{(0)}$ and $k^{(0)}$ of the input state: \ba
d^{(n)}&=&(1-(\cos\phi)^{2n})\,p\,+\,(\cos\phi)^{2n}\,d^{(0)}\,,
\label{itert}\\ k^{(n)}&=& k^{(0)}\,\big(\lambda\,\cos\phi\big)^n
\,. \label{iterk}\ea Thus, whenever $\phi\neq 0$, the iteration of
$T_{\xi}$ yields $d^{(n)}\rightarrow p$ and $k^{(n)}\rightarrow 0$
since $|\lambda|\leq 1$; that is, $\rho^{(n)}\rightarrow \xi$:
almost all the machines of the form (\ref{generalu}) satisfy
Requirement II as well. In conclusion, the family of thermalizing
machines for $h\simeq \si_z$ is composed by the
$U_z(\phi,\theta,\alpha)$ given by (\ref{generalu}) with $\phi\neq
0$ \cite{notecos0}.

{\em Dynamical equivalence of machines.} The dynamics
(\ref{itert}) of the diagonal term $d^{(n)}$, that is the {\em
dissipation}, is determined only by $\phi$. The other parameters
$\theta$ and $\alpha$ enter only the dynamics (\ref{iterk}) of the
off-diagonal term $k^{(n)}$, the {\em decoherence}, through the
complex number $\lambda$ given in (\ref{lambda}). Actually,
$\alpha$ plays a trivial role: it simply redefines at each
iteration the axes $x$ and $y$ in the plane orthogonal to $z$.
Apart from this global rotation, the effect of $\lambda$ can be
visualized as follows: the state of the system undergoes a
rotation of $-\theta$ (resp., $+\theta$) in the $(x,y)$-plane if
it interacts with the state $\ket{0}$ (resp., $\ket{1}$) of the
bath, which happens with probability $p$ (resp., $q$). This
dephasing process contributes only to the decoherence rate. Note
also that $|\lambda|$ is unchanged if one replaces $\theta$ by
$\pi+\theta$. Guided by these considerations, we say that two
machines $U_z(\phi,\theta,\alpha)$ and
$U_z(\phi,\theta+n\pi,\alpha')$ that differ only on the value of
$\alpha$ are {\em dynamically equivalent}. We choose \ba
V_z(\phi,\theta)&\equiv&
U_z(\phi,\theta,0)\,=\,U_z\big(\phi,\theta,\alpha\big)\,\big[u(\alpha)
\otimes u(-\alpha) \big] \label{defv}\ea as representative element
of the class. The $V_z(\phi,\theta)$ are diagonal in the Bell
basis: \ba V_z(\phi,\theta)&=&P_{00} + P_{11} +
e^{i(\theta+\phi)}P_{\Psi^{+}} + e^{i(\theta-\phi)}P_{\Psi^{-}}
\ea with $\ket{\Psi^{\pm}}=\frac{1}{\sqrt{2}}\big(
\ket{01}\pm\ket{10} \big)$. The hamiltonian representation is
easily derived: $V_z(\phi,\theta)=
e^{i\frac{\theta}{2}}\,e^{iH(\phi,\theta)}$ with \ba
H(\phi,\theta)&=&\demi\big[\phi\,(\si_{x}\otimes\si_x+\si_{y}\otimes\si_y)
-\theta\,\si_{z}\otimes\si_z\big]\,. \label{hamphi}\ea Finally
note that we can handle the dissipation and the dephasing
processes separately, since
$V_z(\phi,\theta)=V_z(\phi,0)V_z(0,\theta) =
V_z(0,\theta)V_z(\phi,0)$. By the way, $V_z(\phi,0)$ is a
realization of the two-qubit copying machine proposed by Niu and
Griffiths, that defines Eve's optimal individual attack on the
four-state protocol of quantum crypthography \cite{noteniu}.

{\em The partial swap.}  During the whole construction of the
thermalizing machines, we insisted on the fact that the machine
may depend on the direction $z$ defined by the local hamiltonian
$h$. A natural question is whether any of the machines
$U_z(\phi,\theta,\alpha)$ is actually independent of $z$: such a
machine would thermalize the state of the system for all one-qubit
hamiltonians $h=-E\,\hat{n}\cdot\vec{\si}$. It turns out
\cite{ziman} that there exist a unique machine with this property,
which is $V(\phi,-\phi)$. This machine is a {\em partial swap},
since \ba
V(\phi,-\phi)&=&e^{-i\phi}\,\big(\cos\phi\,\one\,+\,i\sin\phi
\,U_{sw}\big)\, , \label{scalar} \ea where
$U_{sw}=V(\frac{\pi}{2},-\frac{\pi}{2})$ is the swap operation,
i.e. it is the unitary operation whose action is
$\ket{\psi_1}\otimes\ket{\psi_2} \longrightarrow
\ket{\psi_2}\otimes\ket{\psi_1}$ for all
$\ket{\psi_1}\,,\,\ket{\psi_2}\,\in\,\compl^2$. The partial swap
conveys the intuitive idea, that at each collision part of the
information contained in the state of the system is transferred
into the bath. This machine is the cornerstone of the
quantum-information process called {\em homogenization}
\cite{ziman}. For the dissipation, i.e. apart from phase
fluctuations, all thermalizing machines are equivalent to the
partial swap: $V(\phi,\theta)=V(0,\theta+\phi) V(\phi,-\phi)$.

This concludes the characterization of the family of the
thermalizing machines. In the remaining of the paper, we study
their properties, first in terms of thermodynamics, then from the
standpoint of quantum information

{\em Relaxation times.} We'd like to pass from the discrete
dynamics indexed by $n$ to a continuous-time dynamics with
parameter $t$. To perform the limit, we set $n=t/\tau_0$, and we
let the interaction time $\tau_0$ go to zero together with $\phi$
and $\theta$, keeping constant the dissipation rate
$\frac{\phi^2}{\tau_0}=\frac{1}{T_1}$ and the phase fluctuation
rate $\frac{2\theta^2}{\tau_0}=\frac{1}{T_{pf}}$. We have
$(\cos^2\phi)^n\,\approx\,(1-\phi^2)^{\frac{t}{\tau_0}}\rightarrow
e^{-\frac{t}{T_1}}$, and $(|\lambda|\,\cos\phi)^n\,\approx\,
\big[(1-2pq\theta^2)(1-\frac{\phi^2}{2})\big]^{\frac{t}{\tau_0}}\rightarrow
e^{-\frac{t}{T_2}}$ with \ba
\frac{1}{T_2}&=&\frac{1}{2\,T_1}+p\,q\,\frac{1}{T_{pf}}\,=\,\frac{1}{2\,T_1}\,\left(1+4p\,q\,
\lim_{\tau_0\rightarrow 0 }\frac{\theta^2}{\phi^2}\right)\,.
\label{relax}\ea Thus in the continuous-time limit, the processes
of dissipation (\ref{itert}) and decoherence (\ref{iterk}) are
exactly exponential: \ba
d(t)&=&e^{-t/T_1}d(0)+(1-e^{-t/T_1})p\,,\\
|k|(t)&=&e^{-t/T_2}|k|(0)\,. \ea with the relaxation times $T_1$
and $T_2$ defined in the usual way \cite{nmrclass}. For $\theta=0$
or at zero temperature, the bound $T_1\geq \demi T_2$ (see e.g.
\cite{alicki}, p. 120) is saturated.

{\em Fluctuation-dissipation (FD) theorem.} A FD theorem links the
fluctuations at equilibrium and the mechanisms of dissipation.
Usually, this link is derived in a different framework
\cite{weiss}: in particular, a continuous spectral density of the
bath is normally assumed, while here $H_B$ exhibits finite gaps.
Nevertheless, one can define a measurable quantity associated to
fluctuations and dissipation. Consider the following protocol.
First, the system is prepared in the equilibrium state $\xi$ and
is measured in the basis of its eigenstates $P_0$ and $P_1$.
Obviously, the mean values of one-qubit observables $A$ are
unaffected by this measurement. Then we let the system undergo $n$
interactions with the bath qubits: from the state $P_j$ ($j=0,1$)
in which it had been found by the measurement, the system evolves
into the state $\rho_j^{(n)}=T_{\xi}^n[P_j]=
(1-c^{2n})\xi+c^{2n}P_j$ according to (\ref{itert}). By the
definition of equilibrium, $p\rho_0^{(n)}+q\rho_1^{(n)}=\xi$; in
particular, the mean value of $A$ holds unchanged. However, due to
the information gained through the measurement, now we have also
access to the following statistical quantity: \ba
F^{(n)}_A&=&\sqrt{p\big[\mbox{Tr}(\delta_0^{(n)}\,A)\big]^2\,+\,
q\big[\mbox{Tr}(\delta_1^{(n)}\,A)\big]^2}\,, \ea where
$\delta_j^{(n)}=\rho_j^{(n)}-P_j$ is the deviation from the
measured state $P_j$ after $n$ interactions. $F^{(n)}_A$ is a
measure of the fluctuations of $A$; the dissipative element can be
seen through the fact that if $F^{(n)}_A\neq 0$, then the
fluctuations have partly erased the information that we had
obtained through the measurement. Writing
$D^{(n)}=(1-(\cos\phi)^{2n})$ we find \ba
F_A^{(n)}&=&D^{(n)}\,\frac{1}{2\,\cosh(\beta E)}\,
\big|\mbox{Tr}(A\si_z)\big|\,. \label{fluctu} \ea In the
continuous time limit, $D^{(n)}$ is replaced by
$D(t)=(1-e^{-t/T_1})$. This is our FD theorem: the fluctuations
$F$ are proportional to the dissipation $D$ through a function of
the temperature. The fluctuations are absent at zero temperature,
while they are maximal at infinite temperature. Usually one
considers the fluctuations of the one-qubit hamiltonian $h$, in
which case $|\mbox{Tr}(h\si_z)|=2E$ the splitting of the energy
levels.

We proceed now to discuss the link between dissipation and
entanglement under two complementary standpoints (a third approach
is given in Ref. \cite{ziman}). The last equality in (\ref{defv})
means that $U_z(\phi,\theta,\alpha)$ is equivalent to
$V_z(\phi,\theta)$ up to local unitaries (LU); thus we can
restrict to the $V_z(\phi,\theta)$ for the study of entanglement.

{\em Dissipation and Entangling Power.} The entangling power of a
unitary operation $U$ has been given different definitions
\cite{zan,kraus,dur}. Here, we are interested in the creation of
entanglement during the thermalization process. In this context,
the natural definition of the entangling power of a thermalizing
machine $V=V_z(\phi,\theta)$ is \ba
{\cal{E}}[V]&=&\max_{\rho}{\cal{E}}\big( V\rho\otimes\xi
V^{\dagger}\big)\,,\ea with ${\cal{E}}(.)$ a measure of
entanglement. As we said above, for two qubits there exist a
measure of entanglement $\cal{C}$, called {\em concurrence}, that
is computable, basically by ranking the eigenvalues of a $4\times
4$ matrix \cite{woot}. This may be a tedious task on the paper,
but is a trivial one for a computer. Performing the optimization,
we find for $p\geq q$ \ba
{\cal{C}}\big[V_z(\phi,\theta)\big]&=&{\cal{C}}\big(
VP_1\otimes\xi V^{\dagger}\big) \,=\, p\,\sin 2\phi.\ea The
maximal entanglement is thus produced when the input state is the
excited state $\ket{1}$. The phases fluctuations, measured by
$\theta$, don't show up: the entangling power of
$V_z(\phi,\theta)$ depends only on the dissipation, measured by
$\phi$. Moreover, if we want to introduce small fluctuations into
the bath \cite{note2} (and a fortiori in the continuous-time
limit), we must consider small values of $\phi$. In this limit,
the entangling power is increasing with dissipation.

{\em Equivalence under LU.} We have noticed above that
$U_z(\phi,\theta,\alpha)$ is LU-equivalent to $V_z(\phi,\theta)$.
This is a particular case of a general theorem \cite{kraus}
stating that any unitary operation on two qubits is LU-equivalent
to $U_d=e^{iH_d}$, where $H_d=\sum_{i=x,y,z}\mu_i
\si_i\otimes\si_i$. In our case $\mu_x=\mu_y=
\frac{\phi}{2}\in[0,\frac{\pi}{4}]$; and
$\mu_z=-\frac{\theta}{2}$. Since
$V_z\big(\phi,\pi+\theta\big)=\big[\si_z \otimes \si_z
\big]\,V_z\big(\phi,\theta\big)$, one can always choose
$\theta\in[-\frac{\pi}{2},\frac{\pi}{2}]$ within LU-equivalence.
Now, for parameters in these ranges, the $\mu_i$ are uniquely
determined \cite{unique}. Therefore $U_z(\phi,\theta,\alpha)$ is
LU-equivalent to $V_z(\phi',\theta')$ if and only if
$\phi=\phi'\in[0,\frac{\pi}{2}]$ and
$\theta'=\theta\,\mbox{mod}\pi$. In conclusion, two thermalizing
machines are LU-equivalent if and only if they are dynamically
equivalent.

{\em Irreversibility.} We conclude on some general considerations
about irreversibility. The thermalization process that we
described is certainly reversible, since it involves only unitary
operations. Thermalization appears as a consequence of
entanglement, not of measurements. The environment does not play
the role of measuring apparatus, but of "waste basket for
information". In fact, the information encoded in the initial
state of the system $\rho$ is not lost after the thermalization
process, but is encoded in a different way, being spread between
the system and the bath. The initial state of the system can be
reconstructed only if one knows which qubits of the bath have
interacted with the system, and in which order. Without this
knowledge, any attempt of reconstruction of the initial state will
fail \cite{ziman}. Thus, irreversibility arises here as the
interplay of two information processes: (i) the {\em quantum
information} on the initial state of the system is spread between
the system and the bath, still in a reversible way; (ii) the {\em
classical information} about the order of the collisions is lost,
leading to the practical impossibility of running the process
backwards. As an application, one can define "safes" for quantum
information that can be "opened" with classical keys \cite{ziman}.

In summary: we have discussed the family of the thermalizing
machines for two qubits. These unitary operations can be
decomposed into two processes, dissipation and decoherence.
Dissipation is linked to fluctuations --- and this is expected,
although our FD theorem is derived in a different framework than
usual --- and to the entangling power of the machine in the
process. Both dissipation and decoherence are related to
equivalence under local unitaries.

We acknowledge fruitful discussions with Ignacio Cirac,
G\"{u}nther Mahler and Sandu Popescu. Part of this work was
prepared during the ESF Conference on Quantum Information
(Gda\'nsk, 10-18 July 2001). This work was partially supported by
the European Union project EQUIP (IST-1999-11053).

\end{multicols}


\begin{thebibliography}{10}
\bibitem{note2} The second part of Req. II, namely, the fact that
the partial state of the bath must not undergo important changes,
will be met in the physical discussion by requiring that the
interactions are weak.
\bibitem{notecos0} In the case $\cos\phi=1$, we have
$d^{(n)}=d^{(0)}$ for all $n$, but still $k^{(n)}\rightarrow 0$,
unless $|\lambda|=1$, i.e. unless $\theta=0$. So
$\rho^{(n)}\rightarrow dP_0+(1-d)P_1$: the iteration of $T_{\xi}$
does not converge to the state of the bath $\xi$, but is a pure
decoherence process.
\bibitem{unique} This uniqueness is stated in: W. D\"{u}r {\em et
al.}, quant-ph/0112124, that refers for the demonstration to a
paper in preparation. In this paper, the ranges for the parameters
are given in a different way, which is found to be equivalent by
direct inspection.
\end{thebibliography}
\end{document}